\documentclass[twoside,12pt]{article}
\usepackage{graphicx}
\begin{document}

\centerline{\Large \bf Simulating Male Selfish Strategy}
\medskip

\centerline{\Large \bf  in Reproduction Dispute}
\bigskip

\centerline{\bf P.M.C. de Oliveira and S. Moss de Oliveira.}

\bigskip

\noindent Laboratoire PMMH, Ecole Sup\'erieure de Physique et Chimie
Industrielle, 10 rue Vauquelin, F-75231 Paris, France.
\medskip

\noindent
Visiting from Instituto de F\'{\i}sica, Universidade
Federal Fluminense; Av. Litor\^{a}nea s/n, Boa Viagem,
Niter\'{o}i 24210-340, RJ, Brazil; pmco@if.uff.br, 
suzana@if.uff.br.

\bigskip 

\noindent {\bf Abstract}: We introduce into the Penna Model for biological ageing   
one of the possible male mechanisms used to maximize the ability of their 
sperm to compete with sperm from other males. Such a selfish mechanism increases  
the male reproduction success but may decrease the survival probability of 
the whole female population, depending on how it acts. We also find a dynamic phase 
transition induced by the existence of an absorbing state where no selfish 
males survive.
\medskip


 
\section{Introduction}

According to Promislow and Pletcher \cite{pletcher} during the last 20 years there  
have been incredible advances in our undestanding of the evolution of development, 
in molecular evolution, and so on. However, theories of ageing have remained 
relatively restricted to the mutation accumulation theory of Medawar 
\cite{medawar1,medawar2} or to the antagonistic pleiotropy theory of Williams 
\cite{williams}. In their opinion, there are major evolutionary questions 
that could play a very important role in the evolution of senescence. One of these 
questions concerns the conflict between sexes. 

In flies and many other species females mate several times in a short period,  
and store the sperm of different males. Thus, males have evolved mechanisms to 
maximize the ability of their sperm to compete with those of other males, at 
the expenses of the female's fitness. For instance, one protein found in the 
fruit fly male ejaculate has been shown to be similar to a spider neurotoxin 
\cite{wolfiner}.   

Here we introduce into the Penna model for biological ageing \cite{penna}, 
which is based on the mutation accumulation theory, such a kind of male strategy,  
in order to study the consequences on population survival.

\section{The Penna model}

In the asexual version of the Penna model (see \cite{book} for several 
applications)  
the genome of each individual is represented by a computer word
of 32 bits. 
Each bit corresponds to one ``year'' in the individual lifetime, and
consequently each individual can live at most for 32 ``years''. A bit
set to one means that the individual will suffer from the effects of a
deleterious inherited mutation (genetic disease) in that and all following
years. One step of the simulation corresponds to reading one bit of all
genomes. Whenever a new bit of a given genome is read, we increase by one
the individual's age.  The rules for the individual to stay alive are: 1)
The number of inherited diseases (bits set to 1) already accumulated until
its current age must be lower than a threshold $T$, the same for the whole
population. 2) There is a competition for space and food given by
the logistic Verhulst factor $V=1-N(t)/N_{max}$, where $N_{max}$ is the
maximum population size the environment can support and $N(t)$ is the
current population size. We will consider $N_{max}$ five times larger
than the initial population $N(0)$. At each time step and for each
individual a random number between zero and one is generated and compared
with $V$: if it is greater than $V$, the individual dies independently of
its age or genome. 

If the individual survives until a minimum
reproduction age $R$, it generates $b$ offspring in that and all following
years. The offspring genome is a copy of the parent's one, except for some 
randomly chosen deleterious mutations introduced at birth. That is, if a
bit 1 is randomly tossed in the parent's genome, it remains 1 in the
offspring genome; however, if a bit zero is randomly tossed, it is set to
1 in the mutated offspring genome. In this way, for the asexual
reproduction the offspring is always as good as or worse than the parent.

The sexual version of the Penna model was first introduced by
Bernardes \cite{bernardes}, followed by Stauffer et al. \cite{stauffer}, 
who adopted
a slightly different strategy. We are going to describe and use the second
one.  Now individuals are diploids, with their genomes represented 
by two bit-strings that are read in parallel. In order to
count the accumulated number of mutations and compare it with the
threshold $T$, it is necessary to distinguish between recessive and
dominant mutations. A mutation is counted if two bits set to 1 appear at
the same position in both bit-strings (inherited from both parents) or if
it appears in only one of the bit-strings but at a dominant position
(locus). The dominant positions are randomly chosen at the beginning of
the simulation and are the same for all individuals.

The population is now divided into males and females. After
reaching the minimum reproduction age $R$, a female randomly chooses a
male with age also equal to or greater than $R$ to mate. To construct 
one offspring
genome first the two bit-strings of the mother are cut in a random
position (crossing), producing four bit-string pieces. Two complementary
pieces are chosen to form the female gamete (recombination). Finally,
$m_f$ deleterious mutations are randomly introduced. The same process
occurs with the male's genome, producing the male gamete with $m_m$
deleterious mutations. These two resulting bit-strings form the offspring
genome. The sex of the baby is randomly chosen, each with probability 
$50\%$. This whole strategy is repeated $b$ times to produce
the $b$ offspring. The Verhulst killing factor already mentioned works in
the same way as in the asexual reproduction.

\section{Male selfish strategy and results}

To simulate the male ability to increase its own reproductive success 
we attribute to half of the male population a label, to play the role of 
a selfish gene. The offspring of a selfish male may inherit this gene with  
a given probability $p_s$. Non-selfish males produce only non-selfish offspring.   
The selfish male has the advantage that whenever it mates, the female generates 
$b + 1$ offspring (instead of $b$ as the non-selfish males). On the other hand, 
the female pays the bill for this male advantage (as usual). We tested two different 
prices to be paid: 

\noindent case 1) If a female mates with a selfish male she loses forever 
part of her energy to dispute for food and space; That is, her probability to 
die due to the Verhulst factor increases $20 \%$ (or equivalently, for her 
the carrying capacity $N_{max}$ becomes $20 \%$ smaller). Such a punishement 
occurs only once; if she mates again with another selfish male, nothing new 
happens. 

\noindent case 2) Whenever a female mates with a selfish male, her genetic 
death age (programmed since birth as the age where the $T$-th genetic 
disease would appear) is decreased by one, i.e. the female dies one year 
before the expected date. Note that in this case
there is always a price to be paid, that does not depend on the value of any
random number, contrary to the previous one.
 
For both cases we observe that if the selfish gene inheritance probability 
$p_s$ is below a 
given value $p_c$, after many generations no selfish male remains in the 
population. When $p_s$ increases above $p_c$, the percentage of selfish males 
also increases, reaching $100 \%$ for $p_s = 1$. In figure 1 we show how this 
percentage increases with increasing $p_s$, for case 1). Squares correspond to a larger 
population. The curve for case 2) is analogous. 
There is a dynamical phase transition between an absorbing state 
in which selfish males are extinct, and a state where they do survive. The order 
parameter of this transition is the fraction shown in figure 1. 
\bigskip

\bigskip

\bigskip

\begin{figure}[hbt]
\begin{center}
\includegraphics[angle=-90,scale=0.48]{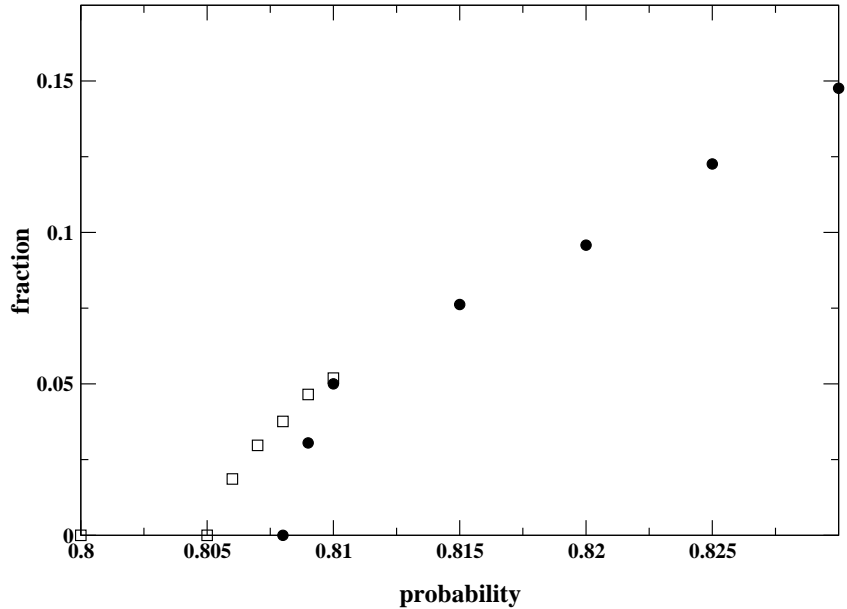}
\end{center}
\caption{Final fraction of selfish males as a function of the inheritance 
probability $p_s$ for case 1.) Common parameters:    
threshold number of bad mutations $T=3$; minimum reproduction age $R=8$; 
birth rate $b=4$; mutation rates $m_m=m_f=1$; results averaged over the 
last 10,000 timesteps. 
Circles: initial population $P(0)=100,000$ and total number of timestpes $t=100,000$; 
Squares: $P(0)=200,000$ and $t=200,000$.}
\end{figure}
 
\begin{figure}[hbt]
\begin{center}
\includegraphics[angle=-90,scale=0.45]{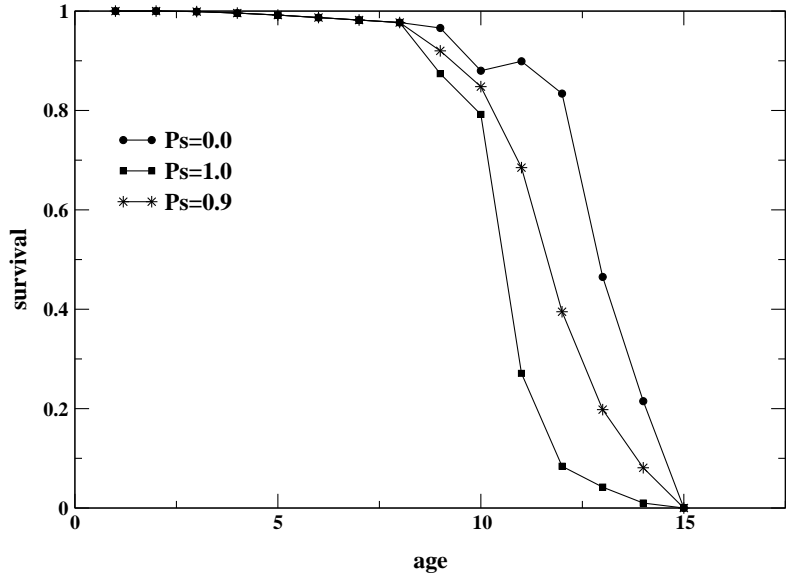}
\end{center}
\caption{Female survival probabilities as a function of age for case 2). The parameters 
are the 
same as those of the larger populations of fig. 1. For any $p_s < p_c$, male 
and female survival probabilities are the same.}
\end{figure}

In both cases the total (already stable) population fluctuates around half of 
the value of the initial population, independently of the $p_s$ value.  
In the first case the male and female population sizes are always the 
same, as well as their survival probabilities as a function of age.
(The survival probability $S(a)$ is defined as the ratio between the number of 
individuals with age $a+1$ and the number of individuals with age $a$.)

However, in the second case the female survival probability decreases 
with increasing values of $p_s$, in comparison to the male survival 
probability (which does not change at all).   
Results are shown in figure 2. 
Also the female population sizes are slightly smaller than the male ones, 
when $p_s \ > p_c$. For an initial population 
of 200,000 individuals (half of each sex), when $p_s < p_c$ both the male and female 
populations stabilize around 87,000. When $p_s=1$, the male population 
stabilizes around 88,000 individuals and the female one, around 84,000.   

\section{Conclusions}

We introduce into the Penna model for biological ageing a male strategy that 
improves its sperm ability to compete with sperm from other males. We 
obtain that depending on how this strategy works, it may decrease the female 
survival probability as a function of age. Considering as an order parameter 
the percentage of males carrying such an ability, we observe a phase transition 
between populations where no such males survive (after many generations) and 
populations where they succeed, depending on the probability of their offspring 
to inherit this ``selfish gene''.

\bigskip

\noindent {\bf Acknowledgements}: To PMMH at ESPCI for the warm hospitality,
to Sorin T\u{a}nase-Nicola for helping us with the computer facilities and
to the Brazilian agency FAPERJ for financial support.

\end{document}